\documentclass{llncs}
\usepackage[latin1]{inputenc}
\usepackage{graphicx}

\begin{document}
\newtheorem{df}{Definition}
\newtheorem{theo}{Theorem}
\newtheorem{lem}{Lemma}
\newtheorem{prop}{Proposition}
\newtheorem{pte}{Property}
\newtheorem{cor}{Corollary}
\newtheorem{rmq}{Remark}
\newtheorem{notation}{Notation}

\title{Imposing edges in Minimum Spanning Tree}

\author{Nicolas Isoart \and  Jean-Charles Régin}
\institute{
Université Côte d'Azur - I3S - CNRS\\
2000, route des Lucioles - Les Algorithmes - Euclide B\\
BP 121 - 06903 Sophia Antipolis Cedex - France\\
isoart@univ-cotedazur.fr, jcregin@gmail.com\\
}
\maketitle

\bibliographystyle{plain}

\begin{abstract}
We are interested in the consequences of imposing edges in $T$ a minimum spanning tree. We prove that the sum of the replacement costs in $T$ of the imposed edges is a lower bounds of the additional costs. More precisely if r-cost$(T,e)$ is the replacement cost of the edge $e$, we prove that if we impose a set $I$ of nontree edges of $T$ then $\sum_{e \in I} $ r-cost$(T,e) \leq$ cost$(T_{e \in I})$, where $I$ is the set of imposed edges and $T_{e \in I}$ a minimum spanning tree containing all the edges of $I$. 
\end{abstract}

\section{Preliminaries}

\subsection{Graph Theory}
A {\bf tree} is a connected and acyclic graph. 
A tree $T=(X',E')$ is a spanning tree of $G=(X,E)$ if $X'=X$ and $E' \subseteq E$. The edges of $E'$ are the {\bf tree edges} of $G$ and the edges of  $E-E'$ are the {\bf nontree edges} of $G$.
A minimum weighted spanning tree (mst) of $G$ is a tree whose sum of the cost of the edges it contains is minimum. 

We recall the Optimality Conditions of a mst:
\begin{theo}\label{theospanningtree}
$ $
\begin{itemize}
\item {\bf [Path Optimality Condition]}
A spanning tree $T$ is a minimum spanning tree if and only if it satisfies the following path optimality conditions: for every nontree edge $\{i,j\}$ of $G$, cost$(\{i,j\}) \geq$ cost$(\{u,v\})$ for every edge $\{u,v\}$ contained in the path in $T$ connecting nodes $i$ and $j$. 
\item {\bf [Cut Optimality Condition]}
A spanning tree $T$ is a minimum spanning tree if and only if it satisfies the following cut optimality conditions: for every tree edge $\{i,j\}$ of $G$, cost$(\{i,j\}) \leq$ cost$(\{u,v\})$ for every edge $\{u,v\}$ contained 
in the cut formed by deleting edge $\{i,j\}$ from $T$.
\end{itemize}
\end{theo}
We will call $\{i,j\}$-tree, a tree which must contain the edge $\{i,j\}$ and denote it by $T_{\{i,j\}}$. 

\begin{pte}\label{msedgetree}
Let $G=(X,E)$ be a graph, $\{i,j\} \in E$ be an edge of $G$.  
We compute a minimum spanning $\{i,j\}$-tree of $G$ by merging first the nodes $i$ and $j$ and then by computing a mst. 
\end{pte}

For the sake of clarity we will consider that $T$ is a minimum spanning tree of $G$. The replacement edge of a nontree edge is defined as follows:
\begin{prop}[\cite{dooms07}]\label{acfilterprop}
Let $\{i,j\}$ be a nontree edge of $G$ that we want to impose, and $r_e$ be the edge that is not imposed with the maximum cost contained in the path in $T$  connecting nodes $i$ and $j$.
Then, the tree $T_{\{i,j\}}$ corresponding to the tree $T$ in which the edge $r_e$ has been replaced by the edge $\{i,j\}$ is a minimum spanning $\{i,j\}$-tree of $G$. 
\end{prop}
{\bf Proof: }
If the edge $\{i,j\}$ is added to the tree then a cycle is created and the Path Optimality Condition implies that the edge of the cycle having the largest cost must be removed. Since a minimum $\{i,j\}$-tree is wanted, the edge that must be removed is $r_e$ because it has the largest cost. Thus a  tree $T_{\{i,j\}}$ is obtained. 
This tree satisfies the Path Optimality Condition for all the nontree edges because $T$ does. $T_{\{i,j\}}$ also satisifies the path optimality condition for $r_e$. \qed
$ $\\
Note that it is possible that an edge has no replacement edge, because it closes a path of implied edges.
In this case, we will consider that the replacement cost of this edge is infinite.

\begin{notation}$ $
\begin{itemize}
\item $P(T,i,j)$ the edges of the simple path from $i$ to $j$ in the minimum spanning tree $T$.
\item r-edge$(T,\{i,j\})$ is the replacement edge of the edge $\{i,j\}$ in the minimum spanning tree $T$.
\item r-cost$(T,\{i,j\})$ is the replacement cost of the edge $\{i,j\}$. It is defined by cost$(\{i,j\})$ - cost$($r-edge$(T,\{i,j\}))$.
\end{itemize}
\end{notation}

\begin{figure}
	\begin{center}
		\includegraphics{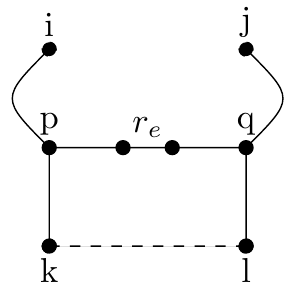}
		\includegraphics{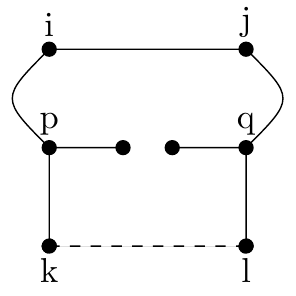}
		\caption{Imposition of the nontree edge $\{i,j\}$ and $\{k,l\}$ in $T$. $T$ is the left graph and $T_{\{i,j\}}$ the right graph.\label{replacementAction} }
	\end{center}
\end{figure}

\begin{prop}\label{propRcost}
$\forall \{k,l\} \not\in T$:
r-cost$(T_{\{i,j\}},k,l) \geq$ r-cost$(T,k,l)$
\end{prop}
{\bf Proof:} two cases must be considered depending on whether  r-edge$(T,\{i,j\})$ belongs to $P(T,k,l)$ or not.

1) r-edge$(T,\{i,j\}) \not\in P(T,k,l)$

In this case,  r-edge$(T_{\{i,j\}},(k,l))=$r-edge$(T,(k,l))$ so the replacement cost is not changed.
  
2) r-edge$(T,\{i,j\}) \in P(T,k,l)$
  
$T_{\{i,j\}}$ is computed by applying the replacement operation from $T$: the edge r-edge$(T,\{i,j\})$ is removed and $\{i,j\}$ is added. Since r-edge$(T,\{i,j\}) \in P(T,k,l)$ then the path from $k$ to $l$ in $T_{\{i,j\}}$ is different from $P(T,k,l)$ because r-edge$(T,\{i,j\}) \not\in T_{\{i,j\}}$. Without loss of generality we assume that $k$ can reach $i$ in $T$ when r-edge$(T,\{i,j\})$ is removed from $T$.  The path $P(T_{\{i,j\}},k,l)$ can be split into three parts: $P(T_{\{i,j\}},k,i)$, $\{i,j\}$ and $P(T_{\{i,j\}},j,l)$. 
The edge $\{i,j\}$ cannot be a replacement edge because it is imposed in the spanning tree. Thus the replacement edge is either in $P(T_{\{i,j\}},k,i)$ or in $P(T_{\{i,j\}},j,l)$.

 $P(T_{\{i,j\}},k,i)$ can also be split into two parts (that can be empty):  $P(T_{\{i,j\}},k,p)$ and  $P(T_{\{i,j\}},p,i)$ where $p$ is the node in $P(T,i,j)$ and in $P(T,k,l)$ whose removal in $T_{\{i,j\}}$ disconnects $k$ and $i$ (See Fig.\ref{replacementAction})\footnote{In fact, there are three possibilities: either there is a path from $i$ to $j$ through $k$, or a path from $k$ to $j$ though $i$, or a fork having $i$ and $k$ as extremities with $p$ in the center and path from $p$ to $j$. We consider only the latter case which is more general.}.
Clearly, we have $\forall \{u,v\} \in P(T_{\{i,j\}},p,i) cost(\{u,v\}) \leq cost($r-edge$(T,i,j))$, because these edges belong to $P(T,i,j)$ and the  replacement edges have the largest cost. Similarly we have $\forall \{u,v\} \in P(T_{\{i,j\}},k,p) cost(\{u,v\}) \leq cost($r-edge$(T,k,l))$. In addition $cost($r-edge$(T,i,j) \leq cost($r-edge$(T,k,l))$ because r-edge$(T,\{i,j\}) \in P(T,k,l)$. 
So, every edge in \\
 $P(T_{\{i,j\}},k,i)$ has a cost that is less than or equal to $cost($r-edge$(T,k,l))$.

A similar reasoning can be applied to $P(T_{\{i,j\}},j,l)$.  $P(T_{\{i,j\}},j,l)$ can also be split into two parts (that can be empty):  $P(T_{\{i,j\}},j,q)$ and  $P(T_{\{i,j\}},q,l)$ where $q$ is the node in $P(T,i,j)$ and in $P(T,k,l)$ whose removal in $T_{\{i,j\}}$ disconnects $j$ and $l$ (See Fig. \ref{replacementAction}). We have $\forall \{u,v\} \in P(T_{\{i,j\}},q,j) cost(\{u,v\}) \leq cost($r-edge$(T,i,j))$, because these edges belong to $P(T,i,j)$ and the  replacement edges have the largest cost. We also have $\forall \{u,v\} \in P(T_{\{i,j\}},l,q) cost(\{u,v\}) \leq cost($r-edge$(T,k,l))$. In addition $cost($r-edge$(T,i,j)) \leq cost($r-edge$(T,k,l))$ because r-edge$(T,\{i,j\}) \in P(T,k,l)$. 
Thus, every edge in  $P(T_{\{i,j\}},j,l)$ has a cost that is less than or equal to $cost($r-edge$(T,k,l))$.

Hence,  $cost($r-edge$(P(T_{\{i,j\}},k,l)) \leq cost$(r-edge$(P(T,k,l))$ so the replacement cost in $T$ is less than or equal to the replacement cost in $T_{\{i,j\}}$ .
\qed
$ $\\
We can now define the wanted proposition:
\begin{prop}
Let $T$ be an mst and $I=\{e_1,e_2,...e_n\}$ a set of nontree edges of $T$. Then,
r-cost$(T_{e_1,e_2,...e_n},k,l) \geq $ r-cost$(T,k,l)$.
\end{prop}

{\bf Proof}
By induction. This is true for one edge. We assume it is true for $n-1$ edges. From Proposition \ref{propRcost} we have  
r-cost$(T_{e_1,e_2,...e_n},k,l) \geq $ r-cost$(T_{e_1,e_2,...e_{n-1}},k,l)$. In addition we have 
r-cost$(T_{e_1,e_2,...e_{n-1}},k,l) \geq $ r-cost$(T,k,l)$. So the proposition holds.\qed
$ $\\
This means that we have the final proposition:
\begin{prop}
Let $T$ be an mst and $I=\{e_1,e_2,...e_n\}$ a set of nontree edges of $T$. Then,
cost$(T_{e_1,e_2,...e_n}) \geq  \sum_{e \in I}$r-cost$(T,e)+$ cost$(T)$.
\end{prop}

\bibliography{jcr}

\end{document}